\documentstyle[11pt,epsf]{article}
\begin{document}
\begin{flushright}
KEK-CP-126 \\
June~2002
\end{flushright}
%\vskip 2cm
\begin{center}
{\Large 
Matching between matrix elements and parton showers
using a Leading-Log subtraction method in NLO-QCD
}
\end{center}
\vskip 0.5cm
\begin{center}
Y. Kurihara \\
\vskip 0.5cm
{\it High Energy Accelerator Research Organization,\\
Oho 1-1, Tsukuba, Ibaraki 305-0801, Japan}
\end{center}
\vskip 0.5cm
\begin{abstract}
A new method to construct event-generators based on
next-to-leading order QCD matrix-elements and leading-logarithmic
parton showers is proposed. 
Matrix elements of loop diagrams as well as those of a tree level 
can be generated using an automatic system. A soft/collinear
singularity is treated using a leading-log subtraction method.
Higher order re-summation of the
soft/collinear correction by the parton shower method is combined with
the NLO matrix-element without any double-counting in
this method.
\end{abstract}
%
% Introduction
%
%\begin{twocolumn}
\section{Introduction}
The standard model(SM) of electromagnetic and strong interactions
has been well established through precision measurements of a large
variety of observables in high-energy experiments\cite{pdg}.
However, there is still one missing part of the SM to be confirmed
by experiments, i.e. the Higgs boson. 
In order to search for those missing parts and also for direct signals
beyond the SM, 
the LHC experiments\cite{lhc} are under construction at CERN.
Though LHC has employed a proton-proton colliding machine due to
a view of the discovery potential for new (heavy) particles, 
a capability for precision measurements must also be 
seriously considered. For extracting physically meaningful information
from a large amount of experimental data, including huge QCD backgrounds,
a deep understanding of the behavior of QCD backgrounds is essentially 
important. 

The prediction power of a lowest calculation of QCD processes is
very limited due to a large coupling constant of the strong
interaction and an ambiguity of the energy scale.
Predictions based on the NLO (or more)
matrix elements combined with some all-order summation is
desirable. 
Much effort to calculate NNLO corrections\cite{nnlo} is being made
for LHC experiments.
Moreover, in order to match with the experimental
requirements for precision measurements, a simple correction
factor (so-called $K$-factor) for the lowest calculation
is clearly not sufficient.
One must construct event generators including NLO matrix elements
with higher order re-summation even for processes with
multi-particle final states. When one tries to do this task,
one may encounter the following difficulties:
\begin{itemize}
\item{the number of diagrams contributing to processes with many particles
in a final state is very large,}
\item{any numerical instability in the treatment of a collinear singularity
must be taken care of,}
\item{a careful treatment of matching between matrix elements
and the re-summation part must be required to avoid double counting, and}
\item{any negative weight of events in some part of phase space is not
avoidable.}
\end{itemize}
In order to overcome these difficulties, lots of
work has already been done\cite{nnlo,gens}. In this report, 
we propose a new method to construct event-generators with
next-to-leading order QCD matrix-elements and leading-logarithmic
parton showers. Our solution for the above problems, except for the last one,
will be given
in this report. A treatment of the
negative weight will be presented elsewhere.

%
% NLO matrix elements
%
\section{NLO matrix elements}
A calculation of the matrix elements of hard-scattering processes
is not such a simple task when multiple partons are produced. 
At the tree level, the {\tt GRACE} system\cite{grace}, 
which is
an automatic system for generating Feynman diagrams and
a {\tt FORTRAN} source-code to evaluate the amplitude, is established.
The system can in principle treat any number of external particles,
and has been used for up to six fermions\cite{tt}
in the final state within a practical CPU time.
In the {\tt GRACE} system, matrix elements are calculated 
numerically using a {\tt CHANEL}\cite{chanel}
library based on a helicity amplitude. {\tt CHANEL} contains
routines to evaluate such things as:
wave-functions/spinors at external states,
interaction vertices, and particle propagator.

For the loop amplitude, effective vertices of two- and three-point
functions have been prepared\cite{guide} and implemented in the system. 
Here, we employed the $\overline{MS}$ scheme\cite{msb} 
for renormalization and dimensional regularization\cite{t-b} 
for treating soft/collinear singularities.
For more than
three-point diagrams, an extended {\tt CHANEL} is prepared. It can evaluate
a fermion current including any number of gamma matrices.
An amplitude of loop diagrams will be numerated based on the fermion
current connected by boson propagators. When a fermion loop is
included in the diagram, numeration will be performed after taking
its trace. A bosonic loop can also be treated by the {\tt CHANEL}.

For a loop integral of four-point functions, general formulae of 
the tensor integral are prepared and implemented in a library.
For more than four-point functions, a reduction formula\cite{guillet}
is used in the system.

%
% Parton shower method
%
\section{Event generation with a parton shower}
\subsection{Conventional method}
The calculations of cross sections for proton-(anti) proton collisions
are usually performed as follows\cite{pythia,herwig}:
\begin{enumerate}
\item{matrix elements of a given process
at a parton level are prepared,}
\item{a probability to observe a parton in a proton at given energy
scale and momentum fraction is
obtained from some parton distribution function(PDF)\cite{pdf},}
\item{a numerical integration is performed in phase-space
with some initial energy distribution in the PDF,}
\item{for initial-state partons,
a (backward) parton shower\cite{ps}
is applied from the given (high) energy scale to a low energy-scale of
partons,} 
\item{for final-state partons, a parton shower is also applied
from a given energy scale to close to the hadron energy-scale,}
\item{some non-perturbative effects according to
string fragmentation or color clustering are taken into account, and}
\item{physical hadrons are formed based on some hadronization model.}
\end{enumerate}
In this method, the parton shower(PS) is used in a supplementary way,
just for generating multi-partons with a finite transverse momentum.

\subsection{An $x$-deterministic PS}
The parton shower
is a method to solve a DGLAP evolution equation\cite{dglap}
using a Monte Carlo method. The PDF's are also obtained by solving
the DGLAP equation with experimental data-fitting.
When an initial distribution of partons is given at some energy scale,
the PS can reproduce a consistent result as the PDF.

The PDF can give the weight of partons when the momentum fraction ($x$) and
the energy scale are given by the users.
On the other hand, in the forward evolution scheme of the PS, 
the momentum fraction of a parton ($x$)
is determined only after evolution takes place. This method is very
inefficient, for instance, for a narrow-resonance production.
In order to cure this inefficiency, we have developed an
$x$-deterministic PS.
In a Monte-Carlo procedure in the PS,
the $Q^2$ evolution is controlled by the Sudakov form-factor, which gives
a non-branching probability of partons, 
and the $x$ determination is done independently
according to a splitting function.
It is not necessary to determine the $x$ value at each branching.
After preforming all branching procedures using the Sudakov form-factor,
the $x$ value at each branching is determined to give a total $x$, being
a given value from outside of the PS. In this PS, each event has
a different weight according to the splitting functions and initial 
distribution of the PDF. 
However, if we 
employ an appropriate transformation of the input random numbers,
we can expect a sufficient efficiency with a weight close to unity.

For singlet partons, we employ an evolution scheme based on
momentum distributions rather than those on particle-number
distributions.
This scheme has been proposed by Tanaka and Munehisa\cite{tm} and
shows numerically stable results with high efficiency.

\begin{figure}[htb]
\centerline{
\epsfysize=15cm
\epsfbox{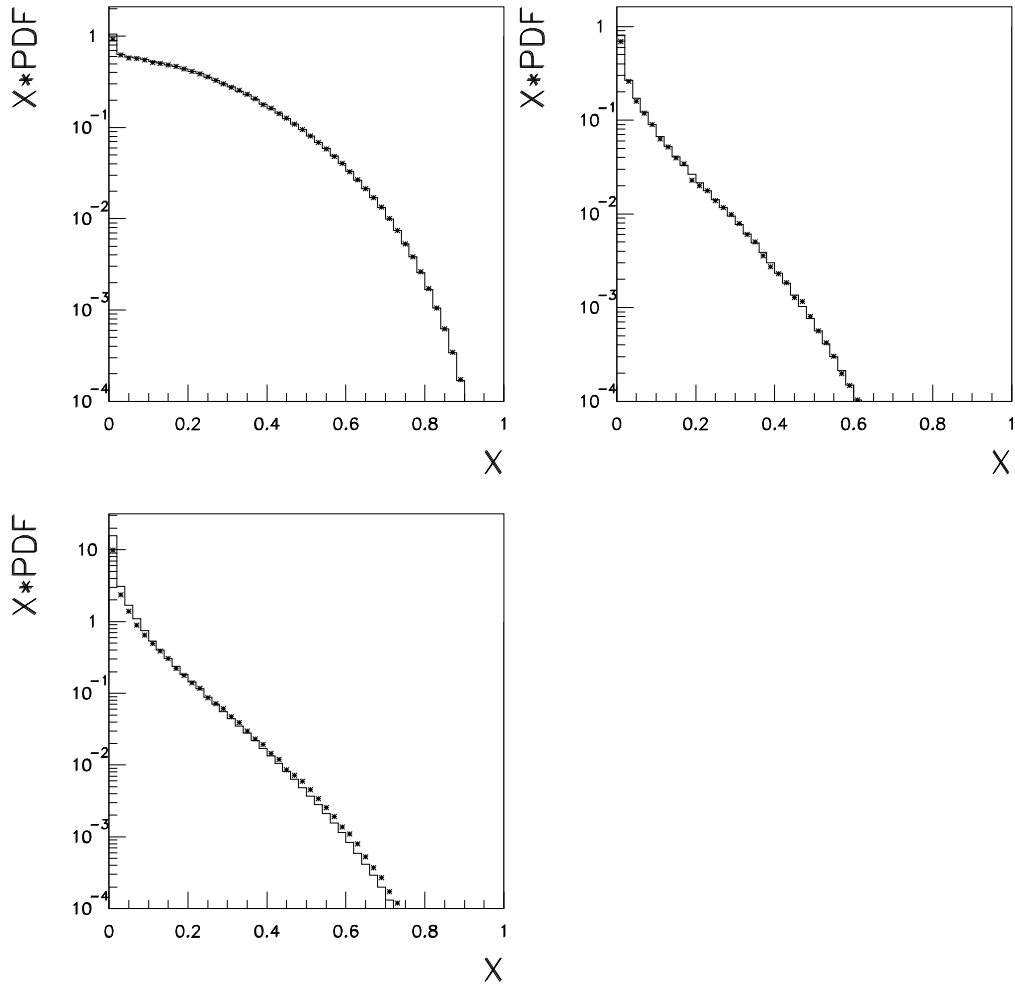}
}
\caption{
Momentum fraction distributions from an x-deterministic parton shower
and Cteq5L parton distribution functions. The distributions of 
$u$-quarks (upper left), ${\bar u}$-quarks (upper right), and
gluons (lower left) are shown.
}
\label{qcdps}
\end{figure}

A numerical test of the $x$-deterministic PS is done by comparing
results with the PDF, {\tt Cteq5L}\cite{cteq}. The initial distributions
of partons are taken from {\tt Cteq5L} at the energy scale of
the $b$-quark mass. Then, after evolution to an energy scale of
100 GeV, the $x$ distributions from the PS are compared to 
those in {\tt Cteq5L}.
One can see in Fig.\ref{qcdps} the PS reproduced 
$x$ distributions in {\tt Cteq5L} for both of singlet and non-singlet
partons.

Our procedure to generate QCD events involves the following three steps,
\renewcommand{\labelenumi}{\theenumi$'$.}:
\begin{enumerate}
\setcounter{enumi}{1}
\item{numerical integration is performed in the phase-space
including $x$ integration,}
\item{the probability for the existence of each parton 
in the proton at some low-energy scale is
obtained from the PDF, and}
\item{the $x$-deterministic PS performs evolution from a low energy scale
to the energy scale of a hard scattering
while generating multi-partons with a finite transverse momentum,}
\end{enumerate}
instead of steps 2, 3, and 4 in the conventional procedure.

%
% Real emission
%
\section{Real emission}
\subsection{Soft/Collinear treatment}
At first, the conventional method of phase-space slicing is used
to treat soft and collinear singularities. It is explained for the
case of an initial-state radiation in this report. 
The final-state radiation can be
treated in a similar way.
Let's consider two colored partons, whose momenta are $p^\mu_1$ and $p^\mu_2$,
scattered into the $N$-body final state of colorless particles 
as a Born process. 
For the radiative
correction of this process, we must treat processes with one additional
colored parton, whose momentum is $k^\mu$, 
emitted in addition to the Born process.
The matrix elements of the real emission processes must be integrated
in ($N$+1)-body phase-space, $\Phi^{(d)}_{N+1}$, in a $d$-dimensional 
space-time.
%\begin{eqnarray}
%d\Phi^{(d)}_{N+1}&=&\frac{d^{(d-1)} {\bf k}}{(2 \pi)^d 2 k^0_i} 
%\prod_{i=1}^{N}\biggl[ \frac{d^{(d-1)} {\bf q}_i}{(2 \pi)^d 2 q^0_i}\biggr]
%(4 \pi)^d \delta^{(d)}(p^\mu_1+p^\mu_2-k^\mu-q^\mu), \nonumber \\
%q^\mu&=&\sum_{i=1}^N q^\mu_i. \nonumber
%\end{eqnarray}
The space-time dimension is set to be
$d=4+2\varepsilon_{IR}$ in this report.
A collinear region in the ($N+1$)-body phase-space 
of the final particles in the $d$-dimensional space-time is defined as
\begin{eqnarray}
\Omega_{coll}&=&\{k^\mu | Q^2_i=-(p_i-k)^2<Q^2_c\}
\subset\Omega_{full}, \nonumber \\
\Omega_{full}&=&\int d\Phi^{(d)}_{N+1}, \nonumber
\end{eqnarray}
where $Q^2_c$ is some cut-off value. It must be sufficiently small 
so as not to be
observed experimentally. A final result must be independent of
this value. 
In the collinear region, matrix elements can be approximated as
\begin{eqnarray}
\biggl|{\cal M}^{(d)}_{N+1}\biggr|^2&=&
\biggl|{\cal M}^{(4)}_{N}\biggl(q\rightarrow \sum_{i=1}^N q_i\biggr)\biggr|^2 
\frac{16\pi}{s \mu^{2\varepsilon_{IR}}}
\biggl(\frac{\alpha_s}{2\pi}\biggr)
\frac{{\tt P}^{\small (1)}(x)}{k_{\bot}^2}\biggl(\frac{1-x}{x}\biggr), \nonumber \\
q^\mu&=&p_1^\mu+p_2^\mu-k^\mu, 
\nonumber
\end{eqnarray}
where ${\tt P}^{\small (1)}(x)$ is a splitting function
for a given parton splitting at a leading-logarithmic order, and
$k_\bot$ is the transverse momentum of the radiated parton and
$\mu$ the energy scale of the splitting.
The CM-energy of the Born process is 
${\hat s}=q^2=s x$, where $s=(p_1+p_2)^2$.
Here, we can set $d$ to be four in the matrix element of
the Born process, since it has no IR singularity.
In the same approximation, the phase space is expressed as
\begin{eqnarray}
d\Phi^{(d)}_{N+1}&=&d\Phi^{(4)}_N\biggl(q\rightarrow \sum_{i=1}^N q_i\biggr)
%%%%%%%%%%%%%%%%%%%%%%%%%%%%%%%%%%%%%%%%%%
%\nonumber \\
%&\times&
%\frac{1}{8\pi^2 \Gamma(1+\varepsilon_{IR})}
%\biggl(\frac{1}{4 \pi x}\biggr)^{\varepsilon_{IR}} k^{1+2\varepsilon_{IR}}_0
%dk_0 (1-\cos^2{\theta})^{\varepsilon_{IR}} d\cos{\theta}
%\nonumber \\
%&=&d\Phi^{(4)}_N\biggl(q\rightarrow \sum_{i=1}^N q_i\biggr)
%%%%%%%%%%%%%%%%%%%%%%%%%%%%%%%%%%%%%%%%%%
\nonumber \\
&\times&
\frac{1}{16\pi^2 \Gamma(1+\varepsilon_{IR})}
\biggl(\frac{k_\bot^{2}}{4 \pi x^2}\biggr)^{\varepsilon_{IR}}
\frac{1}{1-x}
dx dk_\bot^2.
\nonumber
\end{eqnarray}
The total cross section of real-radiation processes in the collinear region
can be obtained as
\begin{eqnarray}
\sigma_{coll}
&=&\frac{1}{(2p^0_1)(2p^0_2)v_{rel}}
\int_{\Omega_{coll}} d\Phi^{(d)}_{N+1} \biggl|{\cal M}^{(d)}_{N+1}\biggr|^2, \nonumber \\
&=&\frac{1}{(2p^0_1)(2p^0_2)v_{rel}}
\biggl[\int_{\Omega_{full}}d\Phi^{(d)}_{N+1}-
\int_{\Omega_{vis}}d\Phi^{(d)}_{N+1}\biggr]
\biggl|{\cal M}^{(d)}_{N+1}\biggr|^2, 
\nonumber
\end{eqnarray}
where $\Omega_{vis}$ means a visible region of the phase space, such as
\begin{eqnarray}
\Omega_{vis}&=&\Omega_{full}-\Omega_{coll}.
\nonumber
\end{eqnarray}
Then, the collinear cross section can be obtained after integration
as
\begin{eqnarray}
\sigma_{coll}
&=&\sigma_{full}-\sigma_{vis}, \nonumber \\
\sigma_{full}&=&\frac{1}{(2p^0_1)(2p^0_2)v_{rel}}
\int_{\Omega_{full}}d\Phi^{(d)}_{N+1}
\biggl|{\cal M}^{(d)}_{N+1}\biggr|^2, \nonumber \\
&=&\sigma_{0}(s) \frac{\alpha_s}{2 \pi} f_{c}
\biggl[\frac{2}{\varepsilon_{IR}^2}+\frac{2L-3}{\varepsilon_{IR}}
-\frac{\pi^2}{2}+L^2\biggr] \nonumber \\
&+&2\int_0^1 dx \sigma_{0}(x s)\biggl[
\phi(x,\varepsilon_{IR})
+\frac{\alpha_s}{2\pi} f_{c} \biggl(
2\frac{(1+x^2)\ln{(1-x)}}{(1-x)_+}
%\nonumber \\ &-&
-\frac{1+x^2}{1-x}\ln{x}               
-x+L\frac{1+x^2}{(1-x)_+}
\biggr)\biggr], \nonumber \\
\sigma_{vis}&=&\frac{1}{(2p^0_1)(2p^0_2)v_{rel}}
\int_{\Omega_{vis}}d\Phi^{(d)}_{N+1}
\biggl|{\cal M}^{(d)}_{N+1}\biggr|^2, \nonumber \\
&=&\int_0^1 dx \sigma_{0}(x s)\frac{\alpha_s}{2\pi}f_{c}
\biggl[2\frac{1+x^2}{1-x}\ln{\biggl(\frac{s}{Q_c^2}(1-x)-1\biggr)}
\Theta\biggl(1-\frac{2Q_c^2}{s}-x\biggr)\biggr],
\nonumber
\end{eqnarray}
where
\begin{eqnarray}
\sigma_0(xs)&=& \frac{1}{x(2p^0_1)(2p^0_2)v_{rel}} \int d\Phi^{(4)}_N
\biggl|{\cal M}^{(4)}_{N}\biggl(q\rightarrow \sum_{i=1}^N q_i\biggr)\biggr
|^2_{q^2=xs} 
\nonumber
\end{eqnarray}
is the Born cross-section at a CM energy of $q^2=xs$ 
in four-dimensional 
space-time, and $L=\ln{(s/\mu^2)}$,
$\mu$ is a factorization energy scale,
and $f_c$ is a color factor of the given branching.
The IR-divergent terms in the collinear cross section can be 
canceled out after summing up terms from virtual corrections.
Even after combining with virtual corrections, 
there still exists the IR-divergent term,
%\begin{eqnarray}
%\phi(x,\varepsilon_{IR})&=&\frac{1}{\varepsilon_{IR}}
%\frac{\alpha_s}{2 \pi} {\tt P}^{\small (1)}(x).
%\nonumber
%\end{eqnarray}
$\phi(x,\varepsilon_{IR})=\frac{1}{\varepsilon_{IR}}
\frac{\alpha_s}{2 \pi} {\tt P}^{\small (1)}(x)$.
This term is thrown away by hand, since it is counted in the PDF or
PS on the initial partons.
The collinear remnant in the visible region ($\sigma_{vis}$) 
gives large {\bf negative} values when the threshold energy ($Q_c^2$) is
sufficiently small. This large negative number will be compensated by the
large {\bf positive} cross section of real-emission processes.
We propose to combine $\sigma_{vis}$ with exact matrix elements of the
real emission processes before the phase-scape integration.
It will be discussed later in this report.

%
% visible jets
%
\subsection{Visible jet cross-section}
The last part of the NLO cross-section calculation 
is a real emission of the additional parton into the visible region
with exact matrix elements. Those matrix elements of ($N$+1)-parton
production at the final state in four-dimensional space-time 
can be automatically generated using
the {\tt GRACE} system. 
Cross sections are obtained by integrating those
matrix elements under the four-dimensional phase-space as
\begin{eqnarray}
\sigma_{exact}&=&\frac{1}{(2p^0_1)(2p^0_2)v_{rel}}
\int_{\Omega_{vis}} d\Phi^{(4)}_{N+1} 
\biggl|{\cal M}^{(4)}_{N+1}\biggr|^2.
\nonumber
\end{eqnarray}
The space-time dimension is set to be four ($\varepsilon_{IR}=0$) hereafter, 
since there is no IR-divergence
in $\Omega_{vis}$. 
When the threshold energy ($Q_c^2$) is set to be sufficiently small, this
cross section can be larger than the Born cross section due to 
the largeness of the coupling constant ($\alpha_s$). 
A parton-level calculation has no problem, except for this largeness 
of the higher order correction. However, if one tries to calculate 
the cross sections of a proton (anti-)proton collision, one has to combine
the matrix elements with the PDF or PS to construct a proton from 
partons. The PDF and PS include leading-log(LL) terms of the initial-state
parton emission. If one combines the matrix element with the PDF or PS very
naively, one cannot avoid double-counting of these LL-terms.
Our proposal to solve this problem is to subtract the LL-terms
from the exact matrix elements as
\begin{eqnarray}
\sigma_{LLsub}&=&\frac{1}{(2p^0_1)(2p^0_2)v_{rel}}
\int_{\Omega_{vis}} d\Phi^{(4)}_{N+1} \biggl[
\biggl|{\cal M}^{(4)}_{N+1}\biggr|^2-
\biggl|{\cal M}^{(4)}_{N}(s x)\biggr|^2 f_{LL}(x,s) \biggr], \nonumber \\
f_{LL}(x,s)&=&16 \pi^2\biggl(\frac{\alpha_s}{2 \pi}\biggr)
\frac{{\tt P}^{\small (1)}(x)}{k_\bot^2}\biggl(\frac{1-x}{x}\biggr).
\nonumber
\end{eqnarray}
The second term of the integrand is the LL-approximation of the
real-emission matrix-elements under the collinear condition.
There is nothing but '$\sigma_{vis}$' in the last subsection.
This 'LL subtraction method' can cure not only the double-counting of
LL-terms in the real-emission part, but also any large cancellation between
the collinear remnant and the real-emission correction.
When the threshold value is set to be sufficiently small,
the integrand in $\sigma_{LLsub}$ is very close to zero, because the
LL-approximation is very precise around the collinear region. Then,
the result of integration in $\sigma_{LLsub}$ is independent of
the threshold value of $Q_c^2$. 

This subtraction may distort the experimental
observables of additional jet distributions. However, a subtracted LL-part
will be recovered after adding the PS applied in the Born process.
The exact matrix element of real-radiation processes gives only
non-logarithmic terms to the visible distributions in the LL-subtraction
method.
The relation between 
the PS and the LL-subtraction method is as follows.
The LL-approximation cross section in $\sigma_{LLsub}$ is
\begin{eqnarray}
\sigma_{LL}&=&\frac{1}{(2p^0_1)(2p^0_2)v_{rel}}
\int_{\Omega_{vis}} d\Phi^{(4)}_{N+1} 
\biggl|{\cal M}^{(4)}_{N}(s x)\biggr|^2 f_{LL}(x,s) \nonumber \\
&=&\int_0^1 dx \sigma_0(x s) {\cal D}^{\small (1)}(x),\\
{\cal D}^{\small (1)}(x)&=& \int_{Q_c^2}^{Q_{max}^2}\frac{dk_\bot^2}
{k_\bot^2} \frac{\alpha_s}{2 \pi} {\tt P}^{\small (1)}(x),
\nonumber
\end{eqnarray}
where $Q_{max}^2$ is the maximum value of the transverse momentum.
${\cal D}^{\small (1)}(x)$ is just the first term of the all-order
re-summation of
the leading logarithmic terms in the PS.
Actually, the Sudakov form factor can be obtained from 
${\cal D}^{\small (1)}(x)$ as:
\begin{eqnarray}
\Pi(Q^2_{max},Q^2_c)&=&
exp\biggl(-\int^{1}_{0} {\cal D}^{\small (1)}(x) dx\biggr).
\nonumber
\end{eqnarray}
However, there is one essential difference: 
the upper bound of $k_\bot^2$ integration.
In the PS, this integration is performed up to the energy scale
determined by the Born process. Then, if
$\sigma_{LL}$ is subtracted in the full phase space of the ($N$+1)-particle
final state, some part of the LL cross section cannot be recovered
from the Born cross-section with the PS. Then, an
appropriate restriction on the phase-space integration must be applied,
such as
\begin{eqnarray}
{\tilde \sigma}_{LLsub}&=&\frac{1}{(2p^0_1)(2p^0_2)v_{rel}}
\int_{\Omega_{vis}} d\Phi^{(4)}_{N+1} 
\biggl[
\biggl|{\cal M}^{(4)}_{N+1}\biggr|^2 -
\biggl|{\cal M}^{(4)}_{N}(s x)\biggr|^2 f_{LL}(x,s)
\Theta(Q_{B}^2-k_\bot^2)
\biggr], 
\nonumber
\end{eqnarray}
where $Q_{B}^2$ is the energy scale of the Born process.

%
% a test of the LL-subtraction method
%
\subsection{A test of the LL-subtraction method}
The LL-subtraction method is numerically tested for a process of
$u {\bar u} \rightarrow \mu^+ \mu^-$ in proton-anti proton collision
at the CM energy of 2 TeV. Here only non-singlet $u$-quark
is used in this test this test  to avoid an additional complexity. Matrix elements
of the Born process and the real radiation process 
($u {\bar u} \rightarrow \mu^+ \mu^- gluon$) are generated by {\tt GRACE}.
Numerical integration is done using a {\tt BASES}\cite{bases} system.
A non-singlet $u$-quark distribution at the energy scale of
4.6 GeV is taken from {\tt CTEQ5L}. The parton evolution
from this energy scale to that of hard-scattering, i.e. an invariant mass
of the muon-pair (${\hat s}_{\mu\mu}$), is done using the $x$-deterministic PS
for both of the Born and radiative processes. An additional cut of
$\sqrt{{\hat s}_{\mu\mu}}>40$ GeV is also applied.

The distributions of transverse momenta of gluons
are shown in Figure~2. For the Born process, the largest $k_\bot$ from
the PS is filled to histograms if it has a $k_\bot$ greater than 1 GeV.
For the radiative process, that from matrix elements is filled 
when it has passed the same cut as mentioned above. If one does not
take care of double-counting of the LL-terms in the matrix elements
and the PS, the distributions of $d\sigma_{exact}/dk_\bot$ show
larger values compared with the PS in the Born process, as shown in the
left histogram of Figure~2.
In order to avoid double-counting, we required that 
$k_\bot$ of the matrix elements be greater than those of any gluons
from the PS in the radiative process. This double-counting
rejection has a good agreement of the $k_\bot$ distribution
around the low-$k_\bot$ region, as shown in the middle histogram.
However, the PS can still emit an insufficient
number of gluons around a very high-$k_\bot$ region.
If the upper bound of the $k_\bot^2$ integration is set to be
${\hat s}_{\mu\mu}$, the distribution from exact matrix elements
shows a good agreement with that from the PS, as shown in right
histogram.

The distributions from ${\tilde \sigma}_{LLsub}$ combined with
those of the PS on the Born process are compared with
those from exact matrix-elements with double-counting rejection.
In both of the gluon energy and the transverse-momentum distributions,
the LL-subtraction method can give consistent results with
those of the exact matrix elements, as shown in Figure~3.

\section{Conclusions}
A new method to construct event-generators based on
next-to-leading order QCD matrix-elements and an $x$-deterministic
parton shower is proposed. 
Matrix elements of loop diagrams as well as those of a tree level
can be generated by the {\tt GRACE} system. A soft/collinear
singularity is treated using the leading-log subtraction method.
It has been demonstrated that the LL-subtraction method can give
good agreement with the exact matrix elements without any
double-counting problem.
%%%%%%%%%%%%%%%%%%%%%%%%%% acknowledgment %%%%%%%%%%%%%%%%%%%%%%%%%%%%%
\vskip 1cm
The author would like to thank Drs. M.~Dobbs, S.~Odaka, Y.~Takaiwa,
T.~Abe, and ATLAS-Japan.

This work was supported in part by the Ministry of Education, Science
and Culture under Grants-in-Aid Nos. 11206203 and 11440083.
%%%%%%%%%%%%%%%%%%%%%%%%%% ref %%%%%%%%%%%%%%%%%%%%%%%%%%%%%%%%%%%%%%%
%\end{twocolumn}

%\newpage
%%%%%%%%%%%%%%%%%%%%%%%%%%
%%%%%%%%%%%%%%%%%%%%%%%%%%
\begin{figure}[htb]
\centerline{
\epsfysize=15cm
\epsfbox{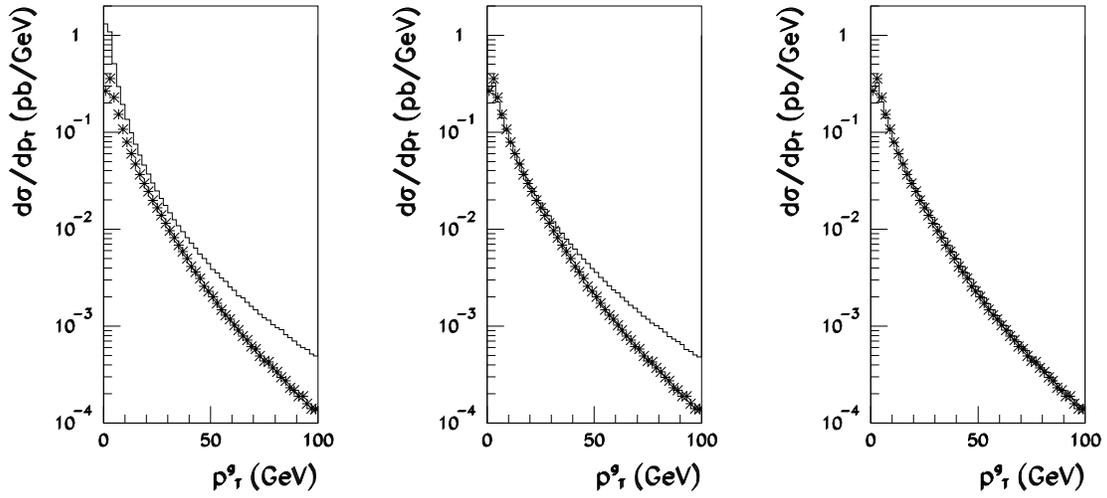}
}
\caption{Transverse momentum distribution of gluons. The distributions
from the PS applied to the Born process are
shown by $*$ commonly in three histograms. Those distributions from
$\sigma_{exact}$ (left), $\sigma_{exact}$ with double-count rejection
(middle), and $\sigma_{exact}$ with double-count rejection and the
$k_\bot^2$ restriction are compared with the PS.
}
\label{fig1}
\end{figure}
%%%%%%%%%%%%%%%%%%%%%%%%%%
\begin{figure}[htb]
\centerline{
\epsfysize=15cm
\epsfbox{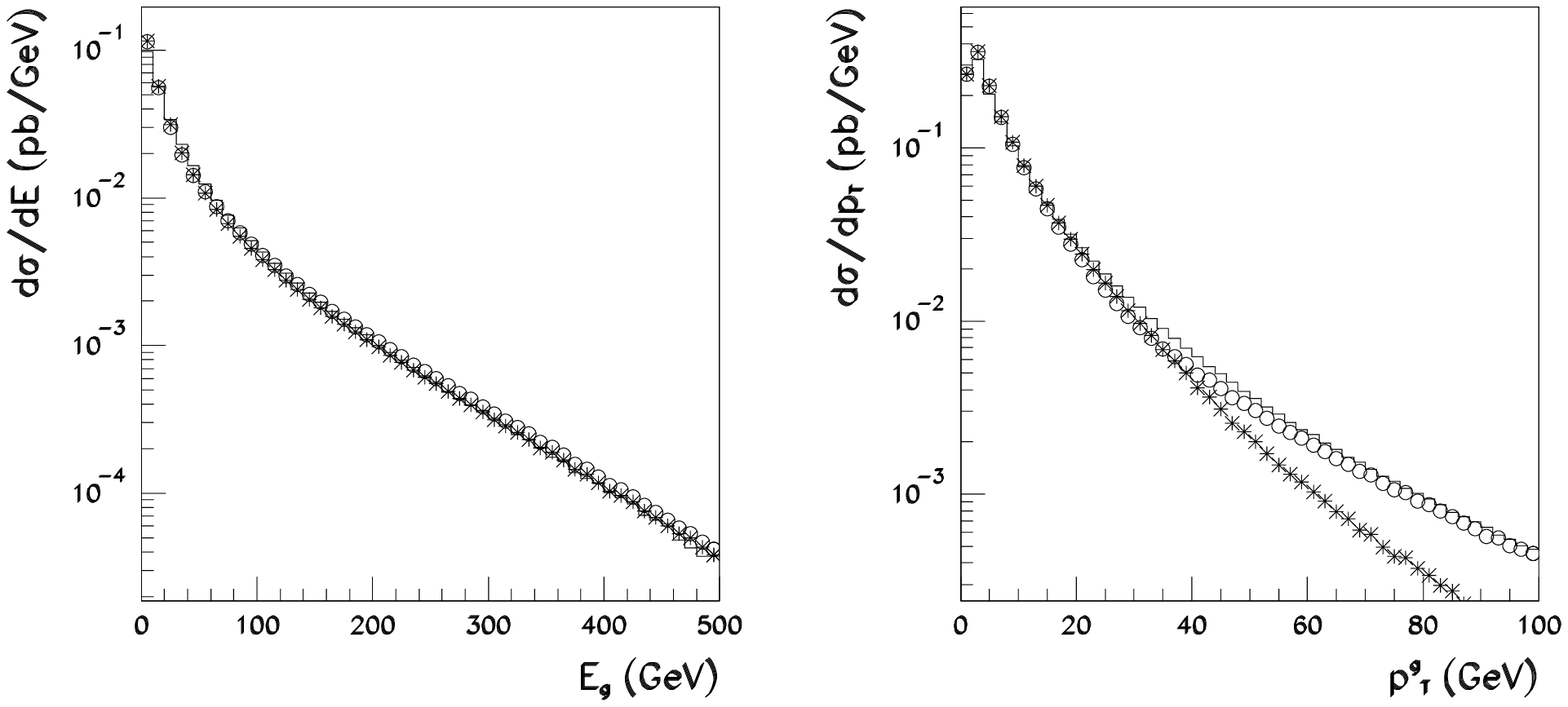}
}
\caption{Energy and transverse momentum distributions of gluons.
The $*$ shows the distributions from the PS on the Born process,
solid histograms from $\sigma_{exact}$ with double-count rejection, 
and circles from ${\tilde \sigma}_{LLsub}$ combined with those
from the PS on the Born process.
}
\label{fig2}
\end{figure}
%%%%%%%%%%%%%%%%%%%%%%%%%%
%\begin{figure}[htb]
%\centerline{
%\epsfysize=10cm
%\epsfbox{respt1s40-3.ps}
%}
%\caption{test
%}
%\label{fig3}
%\end{figure}
%%%%%%%%%%%%%%%%%%%%%%%%%%

\begin{thebibliography}{}
\bibitem{pdg}Particle Data Group, Eur. Phys. J. {\bf C15} (2000) 1
\bibitem{lhc}
ATLAS Technical Proposal, CERN/LHCC/94-43 (1994), \\
CMS Technical Proposal, CERN/LHCC/94-38 (1994). 
\bibitem{nnlo}See ``Physics at TeV Colliders'', Les Houches, France, 
21 May - 1 June 2001,(hep-ph/0204316) and references there in.
\bibitem{gens}
B.~Potter, Phys. Rev. {\bf D63} (2001) 114017, \\
B.~Potter and T.~Schorner, Phys. Lett. {\bf B517} (2001) 86 \\
M.~Dobbs, Phys. Rev. {\bf D64} (2001) 034016, \\
S.~Frixione, R.~B.~Webber, JHEP 0206:029,2002.
\bibitem{grace}
T. Ishikawa, T. Kaneko, K. Kato, S. Kawabata, Y. Shimizu and H. Tanaka,
KEK Report 92-19, 1993, The GRACE manual Ver. 1.0 
{\bf 64}(1991) 149.
\bibitem{tt}
F.~Yuasa, Y.~Kurihara, S.~Kawabata, Phys. Lett. {\bf B414} (1997) 178.
\bibitem{chanel}
H. Tanaka, Comput. Phys. Commun. {\bf 58} (1990) 153 \\
H. Tanaka, T. Kaneko and Y. Shimizu, Comput. Phys. Commun.
\bibitem{guide}
M.~A.~Nowak, M.~Praszalowicz, Ann. of Phys. {\bf 166} (1986) 443.
\bibitem{msb}
A.~Bardeen, A.J.~Buras, D.W.~Duke, T.~Muta, Phys. Rev. {\bf B133} (1964)
1549.
\bibitem{t-b}
G.~'t~Hooft, M.~Veltman, Nucl. Phys. {\bf B44} (1972) 189.
\bibitem{guillet}
Z.~Bern, L.~Dixon, D.A.~Kosower, Phys. Lett. {\bf B302} (1993) 299,\\
J.~Fleisher, F.~Jegerlehner, O.V.~Tarasov, Nucl. Phys. {\bf B566} (2000)
423, \\
T.~Binoth, J.Ph.~Guillet, G.~Heinrich, Nucl. Phys. {\bf B572} (2000) 361.
\bibitem{pythia}
T.~Sj\"ostrand, Comput. Phys. Commun. {\bf 82} (1994) 74.
\bibitem{herwig}
G.~Marchesini, B.R.~Webber, G.~Abbiendi, I.G.~Knowles, M.H.~Seymour, L.~Stanco,
Comput. Phys. Commun. {\bf 67} (1992) 465.
\bibitem{pdf}
GRV, M.~G{\" u}ck, E.~Reya, A.~Vogt, Eur. Phys. J. {\bf C5} (1998) 46, \\
MRS, A.D.~Martin, R.G.~Roberts, W.J.~Stirling, R.S.~Thorne, 
Eur. Phys. J. {\bf C23} (2002) 73, \\
CTEQ Collab., H.L.~Lai et al., Eur. Phys. J. {\bf C12} (2000) 375.
\bibitem{ps}
G.~Marchesini, B.R.~Webber, Nucl. Phys. {\bf B238} (1984) 1,\\
R.~Odorico, Nucl. Phys. {\bf B172} (1980) 157,\\
T.~Sj\"ostrand, Comput. Phys. Commun. {\bf 79} (1994) 503.
\bibitem{dglap}
V.N.~Gribov, L.N.~Lipatov, Sov. J. Nucl. Phys. {\bf 15} (1972) 298,\\
G.~Altarelli, G.~Parisi, Nucl. Phys. {\bf B126} (1977) 298,\\
Y.L.~Dokshitzer, Sov. Phys. JETP {\bf 46} (1977) 641.
\bibitem{tm}
H.~Tanaka, T.~Munehisa, Mod. Phys. Lett. {\bf A13} (1998) 1085.
\bibitem{cteq}
CTEQ Collab., H.L.~Lai et al., Eur. Phys. J. {\bf C12} (2000) 375.
\bibitem{bases}
S.~Kawabata, Comp. Phys. Commun. {\bf 41} (1986) 127;
{\it ibid.,} {\bf 88} (1995) 309.
%\bibitem{catani}
%S.~Catani and M.~H.~Seymour, hep-ph/9605323.
\end{thebibliography}
\end{document}